\documentclass[twocolumn,showpacs,amsmath,amssymb,
superscriptaddress,prb,footinbib]{revtex4}

\usepackage{epsf}
\usepackage{graphicx}
\usepackage{dcolumn}
\usepackage{bm}
\usepackage{pstricks,pst-plot}

\def\({\left(}
\def\){\right)}
\def\[{\left[}
\def\]{\right]}

\begin{document}

\title{Vertically extended Frenkel-Kontorova model: a real space
renormalization group study}%

\author{Javier Rodr\'{\i}guez-Laguna}%
\affiliation{Dto. F\'{\i}sica Te\'orica, Universidad Complutense de
  Madrid, Madrid, Spain.}%
\author{Silvia N. Santalla}%
   \email[]{ssantall@fis.uc3m.es}
   \homepage[]{http://moria.uc3m.es/~noema}
\affiliation{Departamento de F{\'{\i}}sica, Universidad Carlos III
   de Madrid, Legan{\'e}s, Spain.}

\date{\today}

\begin{abstract}
A modification of the Frenkel-Kontorova model is presented in which
particles are allowed to move in the vertical direction. This model is
used to study the formation of islands for a monolayer of 1D
interfaces and the corresponding roughness transition. Both analytical
and numerical approaches are employed, and the numerical algorithm is
based upon real space renormalization group techniques.
\end{abstract}

\pacs{68.35.Ct, 81.15.Hi, 05.10.Cc}

\maketitle

\section{\label{introduccion}Introduction}

The Frenkel-Kontorova (FK) model\cite{BKlibro,ChLub} was originally
proposed by J. Frenkel and T. Kontorova in 1938 to study plastic
deformations and twinning \cite{FK}. It was independently discovered
by F.C. Frank and J.H. van der Merwe in 1949 to study 1D
dislocations\cite{FvdM}. A rigid substrate is modelled by a sinusoidal
potential, and film particles move within it making up a chain linked
with harmonic springs of a given equilibrium length $a_f$. The period
of the sinusoidal potential is assumed to be $a_s=1$. Given the
horizontal positions of the film particles $\{x_i\}_{i=1}^N$, the
system energy is

\begin{equation}
E[x_i]=\sum_{i=1}^N J_s \cos(2\pi x_i) + \sum_{i=1}^{N-1}
J_f (x_{i+1}-x_i-a_f)^2
\label{fk38}
\end{equation}

At zero temperature this model presents a phase diagram in the
variables $J_f/J_s$ and $a_f$ containing many tongues of commensurate
phases separated by infinitesimal gaps of incommensurate
structure\cite{ChouG,Aubry,PA,BM,Ying}. These pathologies are removed
at finite temperature\cite{SAS}, but a continuous phase transition
from a solid to a liquid-like structure remains with $T_c>0$.

The FK model has been extended both to finite-width stripes\cite{BK}
and truly 2D substrates \cite{Bak}. More realistic interatomic
potentials for the film interaction have been employed\cite{MT}, some
of them with a phenomenological ``temperature dependence'' in order to
study structural phase transitions\cite{LZ}.

In this work the FK model is modified so as the film layer may curve,
by allowing particles to displace vertically. Most methods of analysis
of the classical FK model are not suitable in this case. Thus, we have
resorted to a real space renormalization group (RG) numerical
technique, inspired on the analysis of Griffiths and
Chou\cite{ChouG,Chou} and the Density Matrix Renormalization Group
(DMRG) technique\cite{DMRG}. The extension of the DMRG concepts to a
novel area of application may be of interest to practicioners.


Heteroepitaxial surface growth, i.e.: addition on a given crystalline
substrate of particles of a different material with a given lattice
mismatch, has attracted recently considerable technological
interest\cite{Bimberg,DB}. When the lattice parameter is larger in the
film than in the substrate, 3D islands may develop, making up quantum
dots. There are three relevant growth modes: (a) Frank-van der Merwe
(FvdM), consisting only of flat layers; (b) Stranski-Krastanov (SK),
in which 3D islands grow on top of a flat {\em wetting layer} and (c)
Volmer-Weber (VW), in which the 3D islands appear directly on the
substrate.

Many theoretical models have been proposed, either within continuous
elasticity theory\cite{SB} and atomistic models\cite{Bar}. We will
focus on the semi-atomistic FK approach. It has been applied to study
the mechanism of island formation both for 1D\cite{RZ} and 2D
substrates with a realistic interatomic film potential\cite{KTM}. For
a given number of particles, if all monolayers cover completely the
substrate, a FvdM growth mode is considered. If the particles are
distributed in islands, leaving part of the substrate uncovered, a VW
growth is assumed. Whenever a number of covering monolayers (i.e.: a
wetting layer) appears with islands on top of them, the growth mode is
SK. In these works, no vertical deformation of a layer is allowed:
each island consists of a given number of completely flat layers. This
assumption is questionable in the SK case for the reason exposed in
the following paragraph.

Let $a_s$ and $a_f$ be the substrate and film equilibrium lattice
parameters, with $a_f>a_s$. Let us consider a wetting layer composed
of flat monolayers with lattice parameters increasing upwards. The
last monolayer, therefore, is less stressed than the first one. Thus,
it is less likely to develop an islanding mechanism above it. The
formation of a larger film surface has a certain energy cost, which
would not have any counterpart. But this restriction disappears if all
monolayers are allowed to bend. Consider a flat and a curved monolayer
on the same substrate with the same number of atoms: the curved one
must have larger atomic separation. Thus, the wetting layer height
corresponds to the value where the extra energy needed to curve the
film free surface is overcome by the reduction of the internal elastic
energy provided by the effective {\em stretching} of all monolayers.

A first step in this direction is the analysis of the behaviour of a
single monolayer of film which is allowed to curve. This single
monolayer may provide a seed for the future formation of islands when
more monolayers are deposited.

The rest of this article is organized as follows. Section~\ref{modelo}
discusses the vertically extended FK model. Section~\ref{analytical}
provides some analytical insight into the solutions. Section~\ref{RG}
presents the RG-like technique we have employed to get the minimum
energy states and section~\ref{numerical} discusses the numerical
approach, along with the phase diagram for the model.


\section{\label{modelo}Model}

The original FK model splits the energy for each configuration of the
film particles into two parts: the interaction between each deposited
film particle and the substrate is represented using a continuous
sinusoidal substrate potential, and interaction between
nearest--neighbor particles of the film is considered to be harmonic.

This splitting is preserved in our extension, but with some
modifications. First, film particles are allowed to displace in the
direction orthogonal to the substrate. These vertical displacements
should be inhibited by an appropriate increase in the substrate
potential energy. Thus, the sinusoidal substrate potential must be
transformed into a realistic 1+1D potential. Also, the interparticle
film potential must be changed in accordance with the substrate in
order to have the same long distance behaviour. A harmonic film
potential is not realistic: when two neighbour film particles get
sufficiently far away, the interaction should be negligible\cite{MT}.

In our model the $X$ axis shall be parallel to the substrate and the
$Z$ axis shall be orthogonal to it. Particle $i$ has coordinates
$r_i=(x_i,z_i)$ and the configuration energy has the form:

\begin{equation}
E[\{r_i\}]=\sum_{i=1}^N V_s(r_i) + \sum_{i=1}^{N-1} V_f(|r_{i+1}-r_i|)
\label{formal.energy}
\end{equation}

This model shall be referred to as Vertically Extended
Frenkel-Kontorova (VEFK)\cite{poster}. Our target in this work is to
find the ground states, i.e.: the global minimum energy
configurations, and their properties.

\subsection{Film potential}

Anharmonic film potentials for the FK model have been studied for a
long time\cite{MT}. They must fulfill the following conditions: (a) as
we get further away from the minimum, it should be harder to squeeze
particles together than to separate them and (b) asymptotic freedom:
particles should be free when their distance is large. There are many
potentials fulfilling these conditions in the literature, and most
details concerning them are irrelevant for the physics of our
problem. We have chosen the Mie potential\footnote{Other potentials
with similar features, such as the Morse potential or the one proposed
by Markov et al.\cite{MT}, give the same qualitative physics.}.

\begin{equation}   
V_f(d)= \frac{A_0}{\mu_f-\nu_f}\left[\nu_f
  \left(\frac{a_f}{d}\right)^{\mu_f} - 
  \mu_f\left(\frac{a_f}{d}\right)^{\nu_f}\right]
\label{potfilm}
\end{equation}

\noindent where $d\equiv|r_{i+1}-r_i|$ is the actual distance between
neighboring film particles, $a_f$ is their equilibrium distance, $A_0$
is a constant and $\mu_f>\nu_f$. In this work, $\mu_f=12$ and
$\nu_f=6$, which correspond to a Lenard-Jones potential.

We define $K_f$ to be the spring constant of small oscillations around
the equilibrium point of the film potential. In terms of $K_f$, we
find $A_0=K_fa_f^2/\mu_f\nu_f$.

\subsection{Substrate potential}

The FK substrate potential is extended in the vertical ($Z$) direction
in such a way that it reduces to its previous sinusoidal form when
$z=z_{eq}$, a given equilibrium value. Away from it, the potential is
modified so as it is energetically favourable to stay at the
equilibrium height, but not compulsory. The equilibrium points of this
potential must be arranged periodically at points
$(x,z)=(na_s,z_{eq})$, where $n\in {\mathbb Z}$ and $a_s$ is the
substrate lattice parameter, which we shall assume to be $a_s=1$
throughout this article.

\begin{figure}
\includegraphics[scale=0.95]{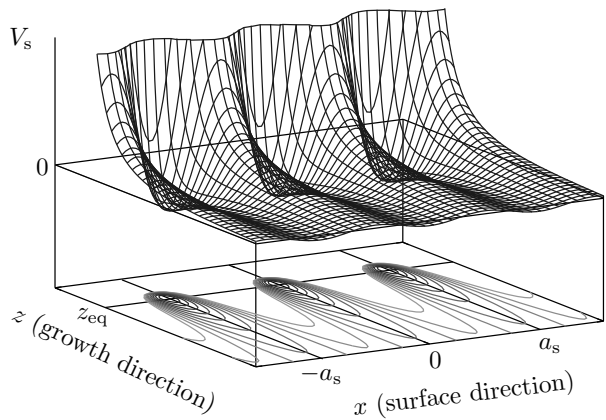}
\caption{\label{fig.substrate}Vertically extended Frenkel--Kontorova
substrate potential.}
\end{figure}

The conditions on the height dependence of the substrate potential are
similar to those for the film: (a) asymptotic freedom in the $Z$-axis,
(b) very high barrier at the substrate height. We propose the
following form

\begin{align}
V_s(x,z)&=\frac{B_0}{\mu_s-\nu_s} \[ \nu_s
\(\frac{z_{eq}}{z}\)^{\mu_s}-\right.\nonumber\\
&-\left. \mu_s \(\frac{z_{eq}}{z}\)^{\nu_s}
\frac{1}{2}\(\cos\(\frac{2\pi x}{a_s}\) +1 \) \]
\label{potential}
\end{align}

\noindent where $B_0$, $\mu_s$ and $\nu_s$ are constants. The $+1$
term by the cosine function makes the potential at $z=z_{eq}$ zero on
average. For $z<z_{eq}$ this average value is positive, and it is
negative, although tending to zero, for $z>z_{eq}$. This ensures that
the ground state is always linked to the substrate.
Figure~\ref{fig.substrate} shows this potential for some typical
values of the parameters. The potential may be seen to be large near
the substrate and zero for large heights. There is an array of minima
with their corresponding corridors, separated by potential barriers
whose height decrease with $z$. The contour view, in the horizontal
plane, depicts some ellipses whose centers correspond to these
minima. We have chosen low values for the exponents: $\mu_s=4$ and
$\nu_s=2$, so as to have deeper corridors and a clearer phase
transition.

\begin{figure}
\includegraphics{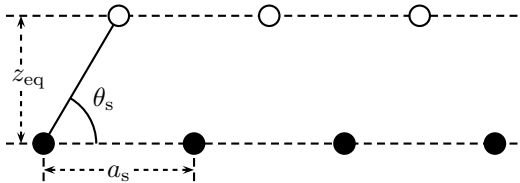}
\caption{\label{fig.equil}Black and white dots represent the substrate
  particles and the substrate potential equilibrium positions
  respectively.}
\end{figure}

The substrate particles creating this potential are arranged at
coordinate points $(x,z)=((n+1/2)a_s,0)$, as is depicted in
figure~\ref{fig.equil}. From the figure it is obvious that
$\theta_s\equiv \arctan(2z_{eq}/a_s)$.

Let us consider a film particle staying close to a minimum of the
substrate potential. From the hessian matrix we may find the second
derivative in the direction of $\theta_s$, towards one of the
substrate particles. If we relate this value to $K_s$, a spring
constant for the substrate-film interaction, we find that

\begin{equation}
B_0= K_s \frac{(\mu_s-\nu_s)(a_s^2+4z_{eq}^2)}
{2 \mu_s (\pi^2 +2\mu_s\nu_s-2\nu_s^2)} 
\label{const.film}
\end{equation}

In the rest of this paper we shall assume $a_s=1$ and
$z_{eq}=0.35$. We define $k\equiv K_f/K_s$. In order to define the
energy scale we assume $K_s+K_f=1$.


\section{Analytical Approach\label{analytical}}

We are going to introduce an Ansatz which provides insight into the
physics of the model. We shall consider all particles to have their
$x$-coordinates in the equilibrium positions of the substrate
potential. Furthermore, let us suppose that a fraction $p$ of them
leave the equilibrium plane $z=z_{eq}$ and reach a new height value
$z_1$. These particles shall be termed {\em runaway particles} and we
shall assume them to be regularly spaced\footnote{The requirement of
regular spacing for the runaway particles is not important within this
Ansatz, i.e.: their distribution may be random as far as there are not
neighbour. This is not true in the general case.}. As a particular
case, if $p=1/2$ we get the {\em zig-zag} state, in which every other
particle leaves the equilibrium plane as a way to relax the film
springs.

\begin{figure}
\rotatebox{270}{\includegraphics[scale=0.34]{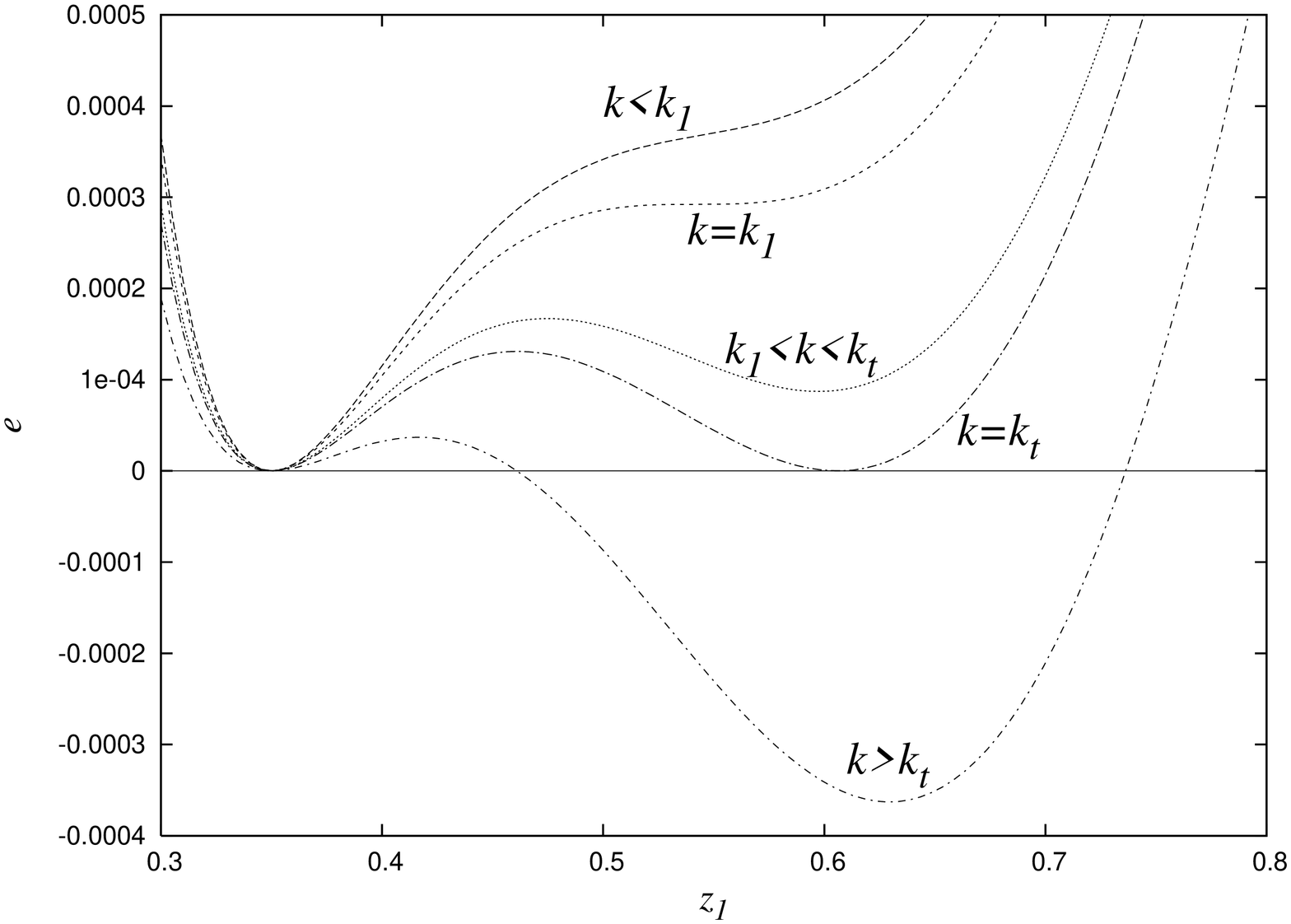}}
\rotatebox{270}{\includegraphics[scale=0.34]{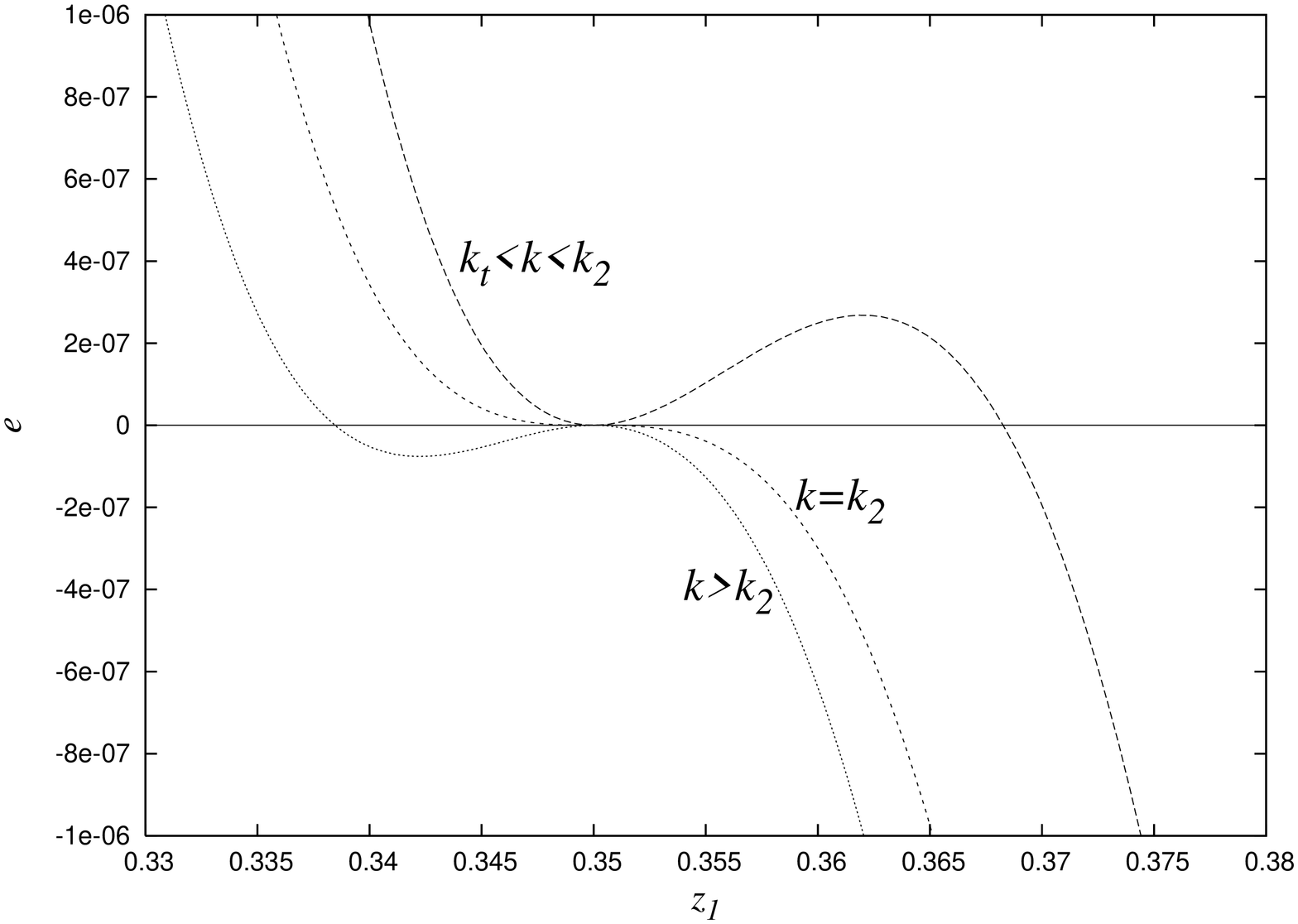}}
\caption{\label{ansatz.E}Function $e(z_1)$ (energy per particle minus
  the value for the flat configuration) within the Ansatz exposed in
  the text. The plots have different values of $k$, but they all take
  $a_f=1.05$ and $p=1/2$. (Upper) The graphs illustrate the behaviour
  of the energy around $k_1$, where a second (rough) minimum appears,
  and $k_t$, where the two minima become equal. (Lower) The graphs
  show the situation around $k_2$, where the first (flat) minimum
  becomes unstable.}
\end{figure}

Let us consider the function $e(z_1)$ to be the energy per particle
minus the value for the flat configuration, for given values of $p$,
$k$ and $a_f$. Figure~\ref{ansatz.E} (upper) shows this function for
five values of $k$, with $a_f=1.05$ and $p=1/2$. The first two cases,
$k<k_1$ and $k=k_1\approx 2.261$, have a single minimum at
$z_1=z_{eq}$. Therefore, the ground state is flat. At $k>k_1$ a second
minimum appears at $z_1>z_{eq}$. The ground state for the third curve,
with $k_1<k<k_t\approx 2.905$, is flat, but there is also a metastable
rough state. At $k=k_t\approx 2.905$ the two minima become equal. This
point marks the {\em rugosity transition}. When $k>k_t$ the ground
state is rough, but the flat state is still metastable for some range
in $k$. Figure~\ref{ansatz.E} (lower) shows that this situation
extends up to $k=k_2\approx 8.128$, when the flat minimum becomes a
saddle point. Beyond that value, for $k>k_2$, the flat state becomes
unstable, and a new metastable state appears, with $z_1<z_{eq}$. The
true ground state is rough and has $z_1>z_{eq}$.

\begin{figure}
\rotatebox{270}{\includegraphics[scale=0.34]{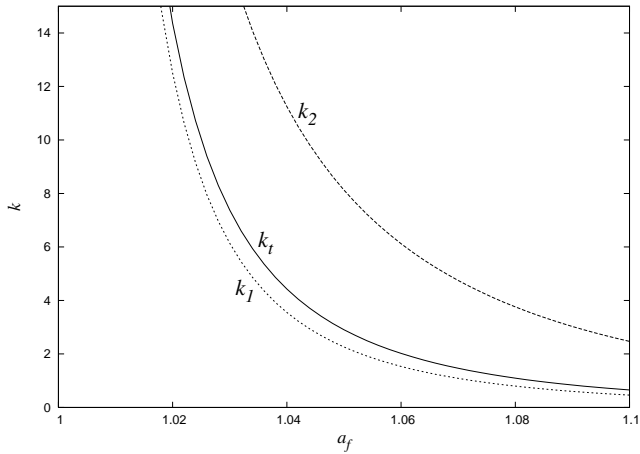}}
\caption{\label{ansatz.phase}Phase transition, $k$ vs $a_f$, for the
  VEFK model within the Ansatz exposed in
  section~\ref{analytical}. The full line labelled as $k_t$ denotes
  the phase transition. Above that line, the ground state is
  rough. Below the $k_1$ curve the ground state is flat. Between $k_1$
  and $k_t$, although the grue ground state is flat, there is also a
  metastable rough state. Between $k_t$ and $k_2$, the true ground
  state is rough, but the flat state is still metastable. Beyond
  $k_2$, the ground state is rough and the flat state is unstable.  }
\end{figure}

Now let us consider the phase diagram, i.e.: for each value of $a_f$,
the value of $k$ for which the two minima of the function $e(z_1)$, at
$z=z_{eq}$ and $z_1>z_{eq}$, take the same energy. This transition
curve $k_t(a_f)$ is depicted in figure \ref{ansatz.phase} with a full
line. The lower curve shows $k_1(a_f)$, the values of $k$ for which a
second local minimum appears at $z>z_{eq}$. Between these two curves,
there is a metastable rough state. The upper curve shows $k_2(a_f)$,
the points where the flat state becomes unstable. Therefore, in the
region between the transition curve and this line, the flat state is
metastable.

The energy $e(z_1)$ is proportional to $p$. Therefore, $p$ plays an
irrelevant role in the determination of the transition point. This
means that, within this Ansatz, the phase transition is not a {\em
truly} collective phenomenon. Since $p=1/2$ is the maximum possible
value of $p$, this Ansatz predicts that beyond the transition point
the rough states will always be {\em zig-zag} states.

The transition curve $k_t(a_f)$ follows with good approximation a
power law $k_t\sim (a_f-1)^\beta$ with two different exponents. For
low values of $a_f$ we have $\beta\approx -1.30\pm 0.02$, and for high
values $\beta\approx -1.94\pm 0.02$. The exponents are not universal:
they depend on the exponents of the substrate and film potentials. We
shall compare these results with the numerical ones.

Although this Ansatz is quite crude, it provides certainly some
insight into the physics of the model. The basic framework, with a
transition curve and the metastability regions, shall remain valid in
the more accurate approach of section \ref{numerical}. We shall see
that $p=1/2$ is not the preferred state after the transition. The
fraction of {\em runaway} particles plays a more relevant role when
the atoms are allowed to move in the $X$-axis.


\section{Renormalization Group Procedure\label{RG}}

\subsection{Formulation}

The main procedure involved by real space RG is to rewrite the system
equations in terms of block variables without altering their form. As
it has been stated, our RG technique is strongly inspired on the
Griffiths and Chou\cite{ChouG} approach and the Density Matrix
Renormalization Group (DMRG)\cite{DMRG}.

\def\ot{\leftarrow} A single particle moves within our 2D substrate
potential, $V_s$, which we shall term $V^{(1)}$ from now on. The
particle stays at position $r_1$, and the system energy is
$V^{(1)}(r_1)$. Let us add a second particle on (e.g.) the right side,
and link them with the ``spring'' $V_f(r_1,r_2)$. The second particle
shall remain fixed at any point $r_2$ and we shall force the first
particle to find its equilibrium position by minimizing
$V^{(1)}(r_1)+V_f(r_1,r_2)$. Thus, we may write $r_1=f_{1\ot 2}(r_2)$
for a known $f_{1\ot 2}$ function. The first particle will be said to
be {\em enslaved} to the second one from now on, and function $f_{1\ot
2}$ shall be called a {\em slaving mapping}.

Now, the total energy of the system may be expressed as:

\begin{align}
V^{(2)}(r_2)\equiv
V^{(1)}(r_1)+V_f(r_1,r_2)+V_s(r_2)\nonumber\\
=V^{(1)}(f_{1\ot 2}(r_2))+V_f(f_{1\ot 2}(r_2),r_2)+V_s(r_2)
\label{rg.first.iteration}
\end{align}

\noindent i.e.: the total energy of the two-particles block depends
only on the position of the second particle. Thus, it may be
considered as an {\em effective} or {\em renormalized} potential. The
minimum value of $V^{(2)}$ is therefore the minimum possible energy
for all the two-particles system. Minimizing for a single particle, we
obtain the minimum for a two-particles chain.

The process may, of course, be iterated until a large chain is
represented this way. If we have $V^{(n-1)}(r_{n-1})$, we may add a
$n$-th particle and find the value of $r_{n-1}$ which minimizes
$V^{(n-1)}(r_{n-1})+V_f(r_{n-1},r_n)$ for each value of $r_n$. This is
again a slaving relation: $r_{n-1}=f_{n-1\ot n}(r_n)$. The $(n-1)$-th
particle is now enslaved to the $n$-th one. Now let us define
$V^{(n)}(r_n)$ as

\begin{align}
V^{(n)}(r_n)\equiv
V^{(n-1)}(r_{n-1})+V_f(r_{n-1},r_n)+V_s(r_n)\nonumber\\
=V^{(n-1)}(f_{n-1\ot n}(r_n))+V_f(f_{n-1\ot n}(r_n),r_n)+V_s(r_n)
\label{rg.iteration}
\end{align}

Using the same argument as before, the minimum value of $V^{(n)}$ is
the minimum possible energy for the whole chain.

\subsection{Reconstruction Procedure\label{reconstruction}}

Let us suppose that we have carried out the previous procedure up to a
given value of $n$ and found the series of potientials
$V^{(1)},\ldots,V^{(n)}$. We may reconstruct the ground state in this
way. The value $r_n^0$ which minimizes $V^{(n)}(r_n)$ is the position
of the $n$-th particle in the ground state. When we have $r_n^0$, we
may find $r_{n-1}$ by minimizing

\begin{equation}
V^{(n-1)}(r_{n-1})+V_f(r_{n-1},r^0_n)
\end{equation}

\noindent Which is equivalent to the use of the enslaving function
$r^0_{n-1}=f_{n-1\ot n}(r^0_n)$. Iterating backwards we can find the
position of the first particle:

\begin{equation}
r^0_1=f_{1\ot 2}(f_{2\ot 3}(\cdots(f_{n-1\ot n}(r^0_n))\cdots))
\end{equation}

\noindent and, thus, obtain the exact solution
$\{r^0_1,\ldots,r^0_n\}$.

Effectively, relation~(\ref{rg.iteration}) may be expanded backwards,
substituting $V^{(n-1)}$ with its definition in terms of $V^{(n-2)}$,
etc. Just as a check, we may evaluate $V^{(n)}(r_n^0)$ and find

\begin{align}
V^{(n)}(r^0_n)&=V_s(r^0_n)+V_f(r^0_{n-1},r^0_n)+ \nonumber\\
&+\sum_{i=1}^{n-1} V_s(r^0_i)+ 
\sum_{i=1}^{n-2} V_f(r^0_i,r^0_{i+1})
\label{reconst.ham}
\end{align}

\noindent Which is, as it was already stated, the energy of the full
chain in the ground state. Thus, the minimization equation for $r_n$
within potential $V^{(n)}$ solves the problem for the $n$-particles
block, which is the essence of a real space RG method.

\subsection{Renormalization Group Flow and Fixed Points} 

We may consider $V^{(i)}$ to be flowing under RG transformations in a
certain functional space. Although the parameter space is infinite
dimensional, this poses no special theoretical difficulties. All
potentials are periodic in $x$ with period $a_s$, since a block
displacement of the full chain of $a_s$ does not affect the energy,
but they need not even be continuous. We will define a fixed point in
this RG flow through the equation:

\begin{equation}
V^{(i)}(r)=V^{(i-1)}(r)+e
\label{fixed.point}
\end{equation}

The parameter $e$ is just the energy per particle in the full
chain. This fixed point equation may be rewritten as an eigenvalue
equation within a minimax algebra\cite{Cunninghame} for a target
potential $R(r_i)$:

\begin{equation}
R(r_i) = V_s(r_i) + \min_{r_{i-1}}[V_f(|r_{i} - r_{i-1}|) +
R(r_{i-1})] + e
\label{fixed.point.2}
\end{equation}

If the flow took place in a finite-dimensional vector space, this
equation would be sure to have a solution\cite{Cunninghame}. In our
problem, despite the absence of a rigorous proof, we have found it to
exist for a wide variety of cases. Solving equation
\ref{fixed.point.2} is a hard problem unless the RG procedure,
starting with $V^{(1)}=V_s$ converges. This convergence has always
taken place in the cases we have tried.

If we are interested only in the thermodynamic limit, we may find the
ground state from the fixed point potential $R$. The procedure is
identical to the one described in the \ref{reconstruction} section,
but using always the same potential. If finite-size effects are
thought to be important, the full series of potentials should be used.


\section{Numerical Approach\label{numerical}}

The most straightforward strategy to obtain the ground state is to
obtain the zero-force equations for each particle, solve them and
choose the solution with lowest energy. The equation for $r_i$ depends
both on $r_{i-1}$ and $r_{i+1}$. This yields a dynamical system which,
for the original FK model, is related to the {\em standard map}. A
minimization algorithm must be used to choose the orbit which
corresponds to the minimum energy. In our case, this strategy is not
appropriate, since it is not possible to solve analytically the $i$-th
zero-force equation for $r_{i+1}$, as it is in the original FK
model. Therefore, the numerical approach shall be based on global
minimization techniques, whose problem is the existence of a complex
landscape of local minima.

In order to obtain a good seed configuration to insert in a global
numerical minimization routine, we have tried several different
algorithms: (a) Simulated annealing\cite{nrc} (b) Take a small chain
and find its ground state; after that, add a site both at the left and
the right borders and repeat the procedure until the desired length is
reached; (c) A technique based on the Puncture Renormalization
Group\cite{prg}: start with any initial configuration, pick up a {\em
block} of sites and minimize its energy while the left and right tails
may only move as two solid blocks; afterwards, the chosen block moves
rightwards and leftwards, {\em sweeping} the system, and (d) The RG
method described in section \ref{RG} whose implementation we are going
to detail.

Method (d) has always provided the lowest energy minima, for reasons
which will be discussed in the following paragraph. The other three
methods give, in many cases, higher energy local minima. Method (b) is
the second best, and its main handicap is that its solutions tend to
have too much left-right symmetry. It is always convenient to optimize
the solution found using a classical minimization algorithm\footnote{A
conjugate gradient search with Polak-Ribiere modification\cite{nrc}.}.

\subsection{Technical Details}

The procedure described in section \ref{RG} may not be carried out
analytically. Our numerical implementation discretizes each $V^{(i)}$
into a regular grid of $N_x\times N_z$ points, typically $100\times
100$ or $100\times 150$ points. The discretization interval of the
potential is an important issue. We have found the election
$[-a_s/2,a_s/2]\times[z_{eq}/2,5z_{eq}/2]$ to be appropriate. The
$X$-interval is motivated by the $a_s$ periodicity of all
potentials. In the $Z$-axis, the interval is designed to include all
runaway particles.

In their seminal work, Griffiths and Chou\cite{ChouG} obtain the {\em
minimum enthalpy} state instead of the minimum energy one. This means
that, in our case, the film potential would be changed into

\begin{equation}
V^*_f(r_1,r_2)\equiv\min_m V_f(|r_1 - r_2 + m\cdot e_x|)
\label{enthalpy}
\end{equation}

\noindent where $e_x$ is the unit vector in the $X$-direction and $m$
is an integer. This change is made so as the search for the minimum in
the construction of $V^{(n+1)}$ from $V^{(n)}$ has a solution in the
$[-a_s,a_s]$ interval of the $X$-axis. Otherwise, the search should be
extended to the whole real line. We have found an alternative approach
to this problem. The search is carried out on {\em two} periods in the
$X$-axis on the left side of the $(n+1)$-th particle. Since $a_f<2a_s$
in all the studied cases, this procedure is sure to find the
solution. Our solution is always optimal within the subspace of
coarse-grained potentials on the given grid.

Our calculations are of two types: {\em infinite systems}, i.e.:
systems in the thermodynamical limit, and {\em finite-size
systems}. The differences between them shall be analyzed in the
following sections.

For each value of $a_f$ and $k=K_f/K_s$ the ground state is found. For
each ground state the following observables are considered: (a) energy
per particle $e$, (b) average height $h\equiv\left<z\right>$ and its
standard deviation $\sigma_h$, known as {\em roughness}, and (c)
effective lattice parameter $\alpha$, defined as the average of
$|x_{i+1}-x_i|$, and its standard deviation $\sigma_\alpha$.

\subsection{Roughness transition: infinite systems
\label{roughness.transition.infinite}}

For any given value of $a_f$, the value $k=K_f/K_s=0$ always results
in a flat configuration, with $h=z_{eq}$ and $\alpha=a_s$. If we
increase $k$, all the observables increase smoothly up to a certain
value $k_t(a_f)$, where an abrupt jump takes place.

The infinite system computation starts with the application of the RG
flow on the substrate potential $V_s$ in order to find the fixed point
potential $R(r)$. Typically, about $10$ or $20$ iterations were enough
to get a maximum error of a part in $10^6$. This fixed point potential
yields a unique enslaving function, $f$. We obtain the point where
$R(r)$ takes its minimum value, $r^0_1$, and iterate the enslaving
function on it so as to reconstruct the full ground state
$\{r^0_1,r^0_2,\ldots,r^0_N\}$ for a large value of $N$. If, for a
given tolerance, there is a value $N_0$ such that if $n>N_0$,
$z_n\approx z_{eq}$, we shall assume that the ground state is flat. We
have chosen a tolerance of $10^{-4}$.

\begin{figure}
\rotatebox{270}{\includegraphics[scale=0.34]{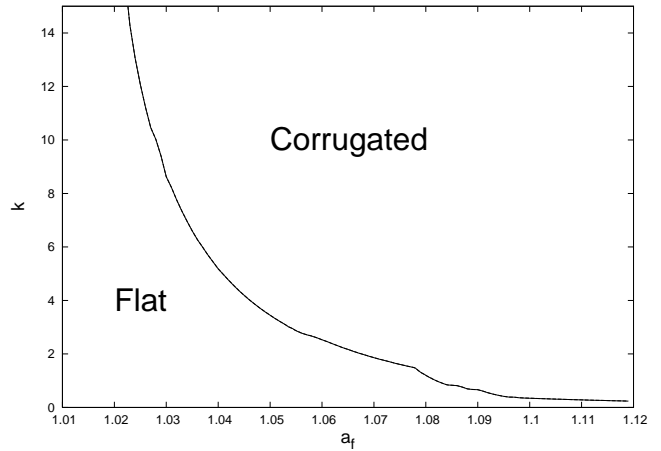}}
\caption{The phase diagram for the VEFK model in the thermodynamical
limit.}
\label{fig.infase}
\end{figure}

The phase transition obtained with this criterion is shown in
figure~\ref{fig.infase}. The first part, up to $a_f\approx 1.075$, may
be fitted to $k_t(a_f)\sim (a_f-1)^\beta$ with $\beta\approx -1.39$,
in reasonable agreement with the value $-1.30\pm 0.02$ obtained for
the analytical Ansatz. For high values of $a_f$ the transition curve
is more rough, and the fit is more difficult. Nonetheless, the
exponent may be seen to increase in absolute value.

The transition always takes place at a higher value of $k$ than in the
analytical Ansatz. This is due to the fact that particles were not
allowed in that case to move in the equilibrium plane. This way to
relax stress delays the transition in practice.

According to the results of section~\ref{analytical}, for most points
just above the transition the ground state is a {\em zig-zag} state,
i.e.: the fraction of {\em runaway particles} is either $p=1/2$ or
$p=1$.

\subsection{Roughness transition: finite-size systems
\label{roughness.transition.finite}}

Finite-size systems present interesting peculiarities. Whenever an
island develops and new film particles are deposited on it, the size
of the island renders the thermodynamical limit irrelevant. Therefore,
finite-size effects might play an important role in practical
applictions. We have mapped and analyzed the roughness transition for
a finite-size system with $N=101$ particles.

In these calculations, the intermediate potentials of the RG flow are
stored to disk so as they may be recovered during the reconstruction
procedure. When convergence is attained, the program stops calculating
any more potentials and assumes the rest to be equal to the last one
in storage. 

The positions of the particles always belong to the $N_x\times
N_z$ grid. Therefore, their precision is low. In the finite-size case
the positions entered a standard minimization algorithm. Usually, this
procedure only took a very short time, since the configuration was
nearly optimal and its deviation was only due to the effect of the
lattice. Checks were periodically made to ensure that the energy per
particle in our results were lower than in the other approaches
described at the beginning of this section.

\begin{figure}
\rotatebox{270}{\includegraphics[scale=0.34]{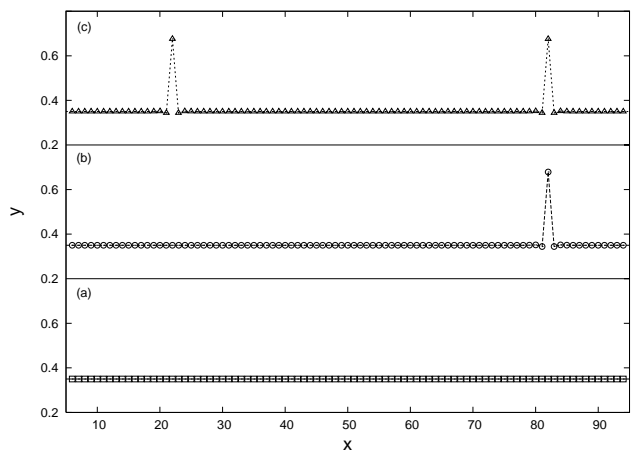}}
\caption{For $101$ particles and $a_f=1.03$, ground states with (a)
$k=3.04670$, (b) $k=3.04674$ and (c) $k=3.04678$.}
\label{fig.tconf}
\end{figure}

Figure~\ref{fig.tconf} shows three ground states with $a_f=1.03$ and
$k=3.04670$, $3.04674$ and $3.04678$ for $N=101$ particles. The first
one is flat. In the second one, a single runaway particle appears. In
the third one, two symetrically placed particles leave the equilibrium
plane $z=z_{eq}$. The values of $\sigma_h$ increase from $7\cdot
10^{-4}$ to $0.032$ and $0.045$ respectively in a very small interval
for $k$. As we increase $k$, new peaks appear steadily, in a slower
way.

\begin{figure}[ht]
\rotatebox{270}{\epsfysize=8cm\epsfxsize=3.5cm\epsfbox{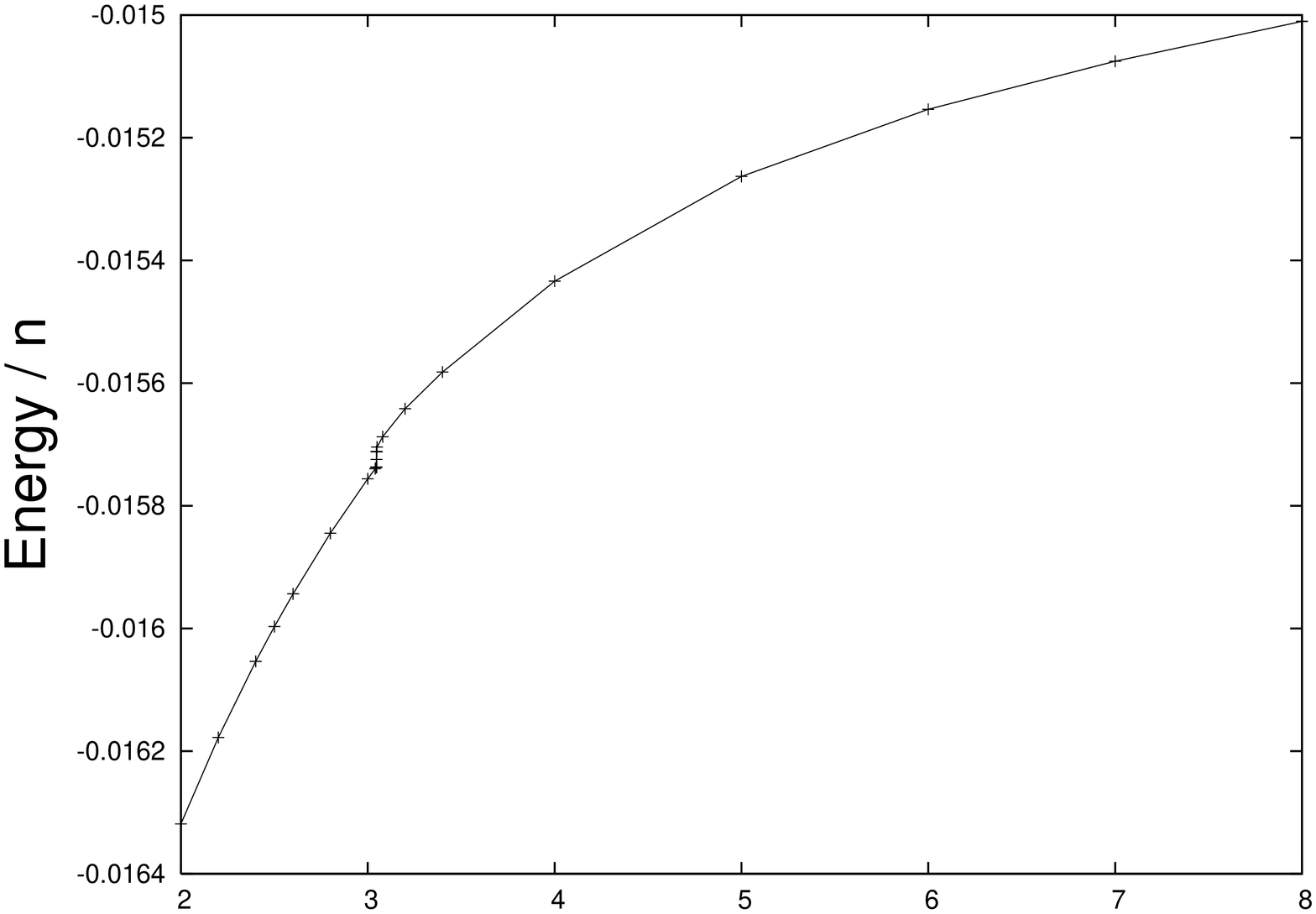}}
\rotatebox{270}{\epsfysize=8cm\epsfxsize=3.5cm\epsfbox{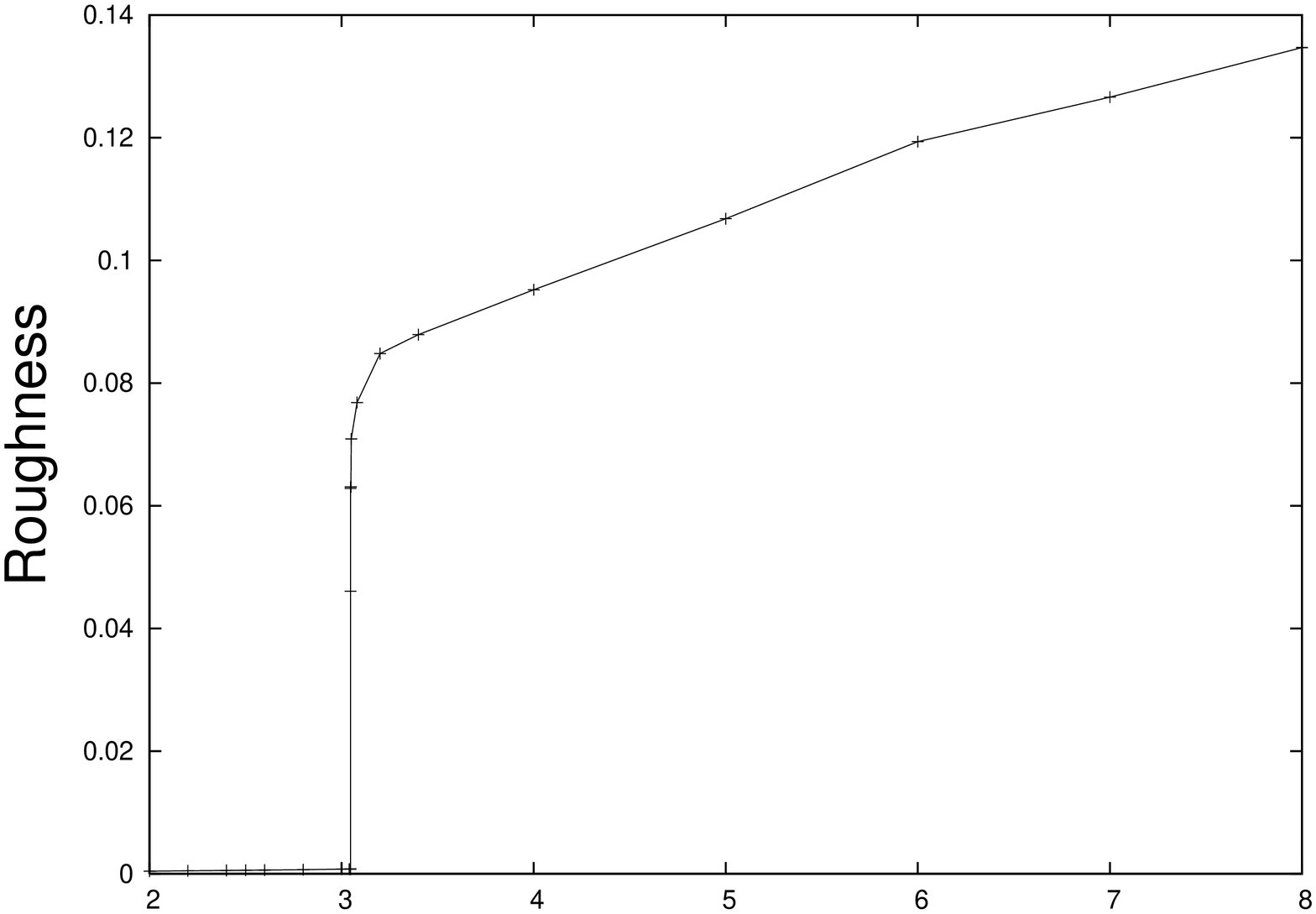}}
\rotatebox{270}{\epsfysize=8cm\epsfxsize=3.5cm\epsfbox{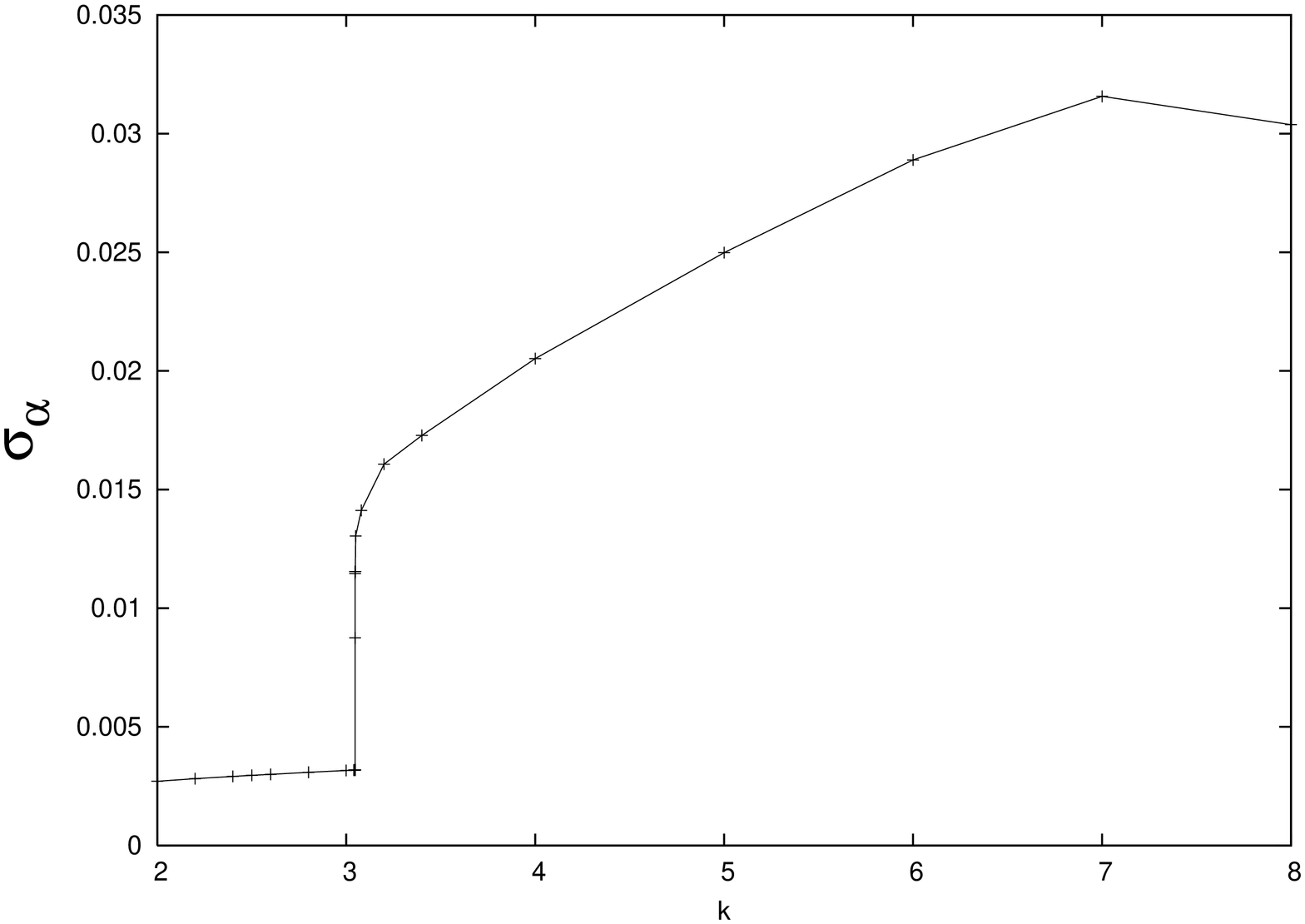}}
\caption{The transition in three observables: (upper) the energy per
  particle, which has a small jump; (middle) the roughness, which has
  a jump of $2$ orders of magnitude; (bottom) the deviation in the
  effective lattice parameter, which also shows a jump. The roughness
  was the guide to find the transition.}
\label{trans.cut}
\end{figure}

Figure~\ref{trans.cut} shows the behaviour of the energy per particle
$e$, roughness $\sigma_h$ and deviation of the effective lattice
parameter $\sigma_\alpha$ for $a_f=1.03$ as we increase $k$. The
transition is clearly seen to coincide for the three observables.

For each value of $k$ and $a_f$ a certain {\em flat domain-size} $D$
may be defined as the minimum distance between runaway particles, or
between runaway particles and the extreme of the chain. Of course, a
ground state is rough whenever $D<N$.

\begin{figure}
\rotatebox{270}{\includegraphics[scale=0.34]{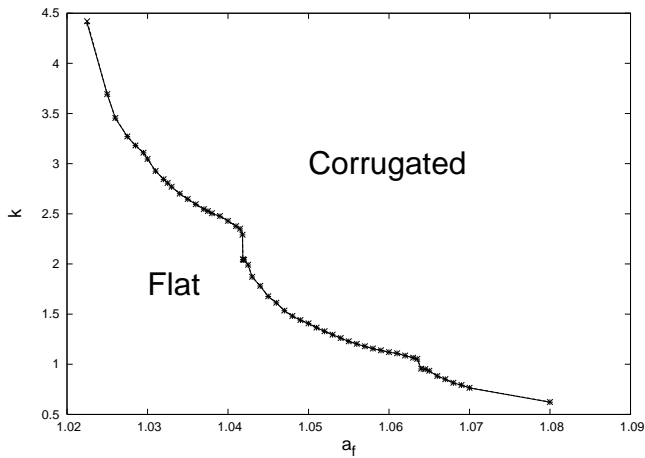}}
\caption{The phase diagram for the VEFK model with $N=101$
particles. Above the line, the ground states are always rough, and
below they are flat. The phase border may be seen to have jumps at
certain positions, as it is explained in the text. 
}
\label{fig.dfase}
\end{figure}

Figure~\ref{fig.dfase} shows the phase transition curve in the $k-a_f$
plane for $N=101$. This curve presents some jumps, e.g. at
$a_f=1.0418$ between $k\approx 2.04$ and $2.28$, which are finite size
effects. At this special jump, the domain size is $D=50$, i.e.: half
the system size. Thus, the first particle to leave the equilibrium
plane is the central one. This behaviour is {\em locked} for a certain
range of $k$ values. The second jump in size is at $a_f=1.064$, and
there the domain size is $D=34$, i.e.: very close to one third of the
total lattice size. This locking is a collective phenomenon which does
not appear in our analytical Ansatz.


This behaviour is clearly different from the one we have found for
infinite systems. Border effects are clearly dominant in the phase
transition for finite systems.


\section{\label{conclusion}Conclusions and Further Work}

A modification of the classical FK model, known as Vertically Extended
Frenkel-Kontorova (VEFK), has been proposed in which particles may
displace vertically, thus allowing the film layer to bend. Physical
pathologies associated with the FK model, such as the sharp
commensurate-incommensurate transition, disappear in this case. This
modified model is used to study a roughness transition which might be
the seed for island formation in surface growth phenomena, both with
analytical and numerical approaches, at $T=0$. The numerical technique
is inspired on the effective potentials method of Chou and
Griffiths\cite{ChouG} and on real space renormalization group methods,
specially on DMRG. The analytical approach shows both a transition
curve and two broad metastability regions.

This roughness transition presents some general features: (a) it
starts with some runaway particles which leave the equilibrium plane,
(b) the size of the flat regions (domains) near the transition point
depends on the parameters $a_f$ and $k$, and is rather sensitive to
finite size effects.

As discussed at the introduction, rough configurations may be the
origin of domains which make up the characteristic islands in SK and
VW growth modes. The effect of many film monolayers is, of course, of
great importance, and has been left for further work. A second
monolayer may have a qualitative effect on the transition.

All the surface growth phenomena are, of course, 2+1D. The dimensional
extension of our model involves some numerical difficulties, since
there does not exist a straightforward extension of the RG
approach. The performance of DMRG-like techniques decrease when the
particle linking graph does not have a tree topology. Extensions of
the DMRG algorithm to 2D and 3D problems are a field of active
research\cite{prg,leiden} and this problem might well pose a possible
application for them.

This VEFK model may be studied at finite temperature with the transfer
matrix formalism. The usual arguments against 1D phase transitions do
not apply in this case, since the substrate potential imposes a long
range order. Therefore, the roughness transition is likely to survive
up to a finite value of $T$. A positive result in this direction is
the treatment of a finite temperature transition for FK.

In physical terms, there should be a second {\em flatness transition}
at even higher values of $k=K_f/K_s$, for which the film gets flat
again, with the effective lattice parameter $\alpha=a_f$. In our
numerical analyses this second transition does not appear. The reason
is the following: at very high values of $k$ the substrate potential
is almost inexistent, and the reference plane for the film disappears
in practice. All configurations in which each particle is at a
distance $a_f$ from its neighbours, even if it is highly curved, have
nearly the same energy.

Some relevant features, such as a slowly changing curvature or the
second flatness transition, would be met if a three-body angular
potential was included. This extension is physically plausible and may
lead to a richer phase diagram.


\begin{acknowledgments}

We have benefitted from very useful discussions with A. Degenhard,
R. Cuerno and M.A. Mart\'{\i}n-Delgado.

\end{acknowledgments}


\end{document}